\documentclass{PoS}
\usepackage{amsmath}
\usepackage{amssymb}
\usepackage{fontenc}
\usepackage{times}
\usepackage{mathptmx}
\usepackage{graphicx}

\ShortTitle{Hadrons as Holograms}

\title{Hadrons as Holograms}
\author{\speaker{Hilmar Forkel} \thanks{%
Supported by the Deutsche Forschungsgemeinschaft.} \\
Institut f\"{u}r Physik, Humboldt-Universit\"{a}t zu Berlin, D-12489 Berlin,
Germany\\
E-mail: forkel@physik.hu-berlin.de}

\abstract{We review our recent work on four topics in strong-interaction 
physics from the perspective of the gauge/gravity correspondence. In 
particular, we discuss
(i) the construction of the ``metric soft wall'' dual for holographic QCD 
which reproduces the observed linear square-mass trajectories of radially 
and orbitally excited (light-quark) hadrons, (ii) an extension of the 
metric soft wall which encodes diquark correlations holographically 
and additionally leads to an excellent description of the empirical 
nucleon excitation spectrum, 
(iii) an AdS/QCD dual that emerges as a new solution of 5d Einstein-dilaton 
gravity with a specifically derived potential and realizes the area law of the 
Wilson loop and (approximately) linear meson trajectories dynamically, and 
finally (iv) the calculation of glueball correlation functions and decay 
constants in both the hard- and dilaton soft-wall gravity duals, as well 
as a quantitative analysis of their QCD-relevant physics 
content.}
\FullConference{International Workshop on QCD Greens Functions, Confinement and
Phenomenology\\
                 September 7-11 , 2009\\
                 ECT Trento, Italy}

\begin{document}
\section{Introduction}

The gauge/string correspondence \cite{aha00} continues to supply new and
exciting perspectives for nonperturbative QCD. By holographically relating
strongly coupled gauge theories to physically equivalent but weakly coupled
string theories, it has provided new analytical tools which promise to 
eventually describe large-$N_{c}$ QCD in terms of a dual, classical string
dynamics \cite{revs}. In order to persue the long-term goal of determining
this dynamics, current bottom-up approaches, often referred to as
\textquotedblleft AdS/QCD\textquotedblright, construct approximate
holographic duals by incrementally encoding known QCD properties 
(experimental and lattice data, low-energy theorems, the operator product
expansion etc.) into 5d gravity backgrounds. After implementing the most
fundamental features, i.e. conformal symmetry breaking, mass gap, quark
confinement etc., one then increasingly incorporates information from the
hadron spectrum and from more detailed amplitudes, and sets out to find the
gravitational dynamics which generates the obtained background as a solution.

The foundational ingredient of any AdS/QCD dual is the metric of its bulk 
spacetime. This geometry contains a five-dimensional, 
non-compact part that takes the generic form of an IR-deformed anti--de 
Sitter space $\mathrm{AdS}%
_{5}\left( R\right) $ of curvature radius $R$~\cite{pol02}, 
\begin{equation}
ds^{2}=g_{MN}^{\left( \text{AAdS}_{5}\right) }dX^{M}dX^{N}=e^{2A\left(
z\right) }\frac{R^{2}}{z^{2}}\left( \eta _{\mu \nu }dx^{\mu }dx^{\nu
}-dz^{2}\right),  \label{metric}
\end{equation}%
where $\eta _{\mu \nu }$ is the four-dimensional Minkowski metric of the 3+1
dimensional boundary on which the gauge theory is defined. For an
UV-conformal gauge theory like QCD, the metric has to approach AdS$_{5}$
near the boundary. Since $A\neq 0$ breaks conformal invariance explicitly,
this amounts to requiring that $A(z)\rightarrow 0 $ as $z\rightarrow 0$. 
A minimal way of implementing the most crucial IR effects, in particular
conformal symmetry breaking and linear quark confinement, is to impose a
boundary condition on the string modes at the IR brane of the
\textquotedblleft hard-wall\textquotedblright\ metric \cite{pol02} 
\begin{equation}
e^{2A_{\text{hw}}\left( z\right) }=\theta \left( z_{m}-z\right) ,\,\ \ \ \ \
\ z_{m}=\Lambda _{\text{QCD}}^{-1}  \label{hw}
\end{equation}%
where $z_{m}$ acts as an infrared cutoff on the fifth dimension and
generates both the mass gap and discrete hadron spectra. The rather drastic
hard-wall approximation underlied all of the first bottom-up duals and
describes a surprising amount of hadron phenomenology \cite{pol02,hwph}.
Hence it provides a useful benchmark for the development of improved 
holographic
duals. The main limitations of the current generation of
AdS/QCD gravity duals are discussed e.g. in Refs. \cite{crit,for08}.

\section{The \textquotedblleft metric soft-wall\textquotedblright\ dual and
linear baryon trajectories}

\label{msw}

One of the most prominent and pervasive patterns in the known hadron
spectrum consists of linear Regge-type trajectories 
\begin{equation}
M^{2}=M_{0}^{2}+W\left( N+L\right)  \label{nltraj}
\end{equation}%
with approximately universal slopes $W\sim 1.1$ GeV$^{2}$ (for the
light-quark mesons \emph{and} baryons)~\cite{slope} on which the square
masses $M^{2}$ of excited states organize themselves in Chew-Frautschi
plots, i.e. as a function of both angular momentum $L$ (or alternatively
total spin $J$) and radial excitation level $N$. The QCD-based understanding
of these trajectories and their relation to linear quark confinement remains
one of the pre-eminent challenges of strong-interaction physics.

A serious limitation of\ holographic duals based on the hard wall (\ref{hw})
is that they predict quadratic instead of linear square-mass trajectories as
a function of $J$, $L$ and $N$ (in the gravity approximation) \cite{hwph}.
The first proposal for correcting this shortcoming, the \textquotedblleft 
\emph{dilaton soft wall}\textquotedblright\ dual \cite{kar06}, generates
linear Regge trajectories $m_{N,J}^{2}\sim N+J$ only in the meson but not in
the baryon sector. Baryon trajectories are similarly pronounced in the
experimental data~\cite{slope}, however, which led us to construct the
\textquotedblleft \emph{metric soft wall}\textquotedblright\ \cite{for07},
the so far only AdS/QCD dual which predicts linear trajectories in the
baryon sector as well. It further shows that (and partially explains why)
universal-slope trajectories (\ref{nltraj}) can be encoded solely into 
IR deformations $A\left( z\right) $.

The metric soft-wall dual is constructed with the help of those string mode
fluctuations in the general geometry (\ref{metric}) which are dual to the
hadronic states under consideration. Casting the wave equations for the
\textquotedblleft radial\textquotedblright\ components of the
(normalizable)\ dual bulk modes into the form of Sturm-Liouville eigenvalue
problems, one has 
\begin{equation}
\left[ -\partial _{z}^{2}+V_{\text{M}}\left( z\right) \right] \varphi _{%
\text{M}}\left( z\right) =M_{\text{M}}^{2}\varphi _{\text{M}}\left( z\right)
\label{eveqm}
\end{equation}%
for the modes dual to spin-0 (M = S) and spin-1 (M = V) mesons as well as 
\begin{equation}
\left[ -\partial _{z}^{2}+V_{\text{B},\pm }\left( z\right) \right] \psi
_{\pm }\left( z\right) =M_{\text{B}}^{2}\psi _{\pm }\left( z\right)
\label{eveqb}
\end{equation}%
from the iterated equation for the chirally decomposed Dirac field 
\begin{equation}
\Psi \left( x,z\right) =\int \frac{d^{4}k}{\left( 2\pi \right) ^{4}}e^{-ikx}%
\left[ \psi _{+}^{\left( k\right) }\left( z\right) P_{+}+\psi
_{-}^{\left( k\right) }\left( z\right) P_{-}\right] \hat{\Psi}^{\left(
4\right) }\left( k\right)  \label{5ddir}
\end{equation}%
($P_{\pm }\equiv \left( 1\pm \gamma ^{5}\right) /2$) dual to spin-1/2
baryons (where $\hat{\Psi}^{\left( 4\right) }$ solves the 4d boundary Dirac
equation) and similarly for 3/2 baryons \cite{hwph}. The eigenvalues $M_{%
\text{M},\text{B}}^{2}$ generate the mass spectra of the gauge theory, and
the potentials are 
\begin{eqnarray}
V_{\text{S}}\left( z\right) &=&\frac{3}{2}\left[ A^{\prime \prime }+\frac{3}{%
2}A^{\prime 2}-3\frac{A^{\prime }}{z}+\frac{5}{2}\frac{1}{z^{2}}\right]
+m_{5,S}^{2}R^{2}\frac{e^{2A}}{z^{2}},  \label{vma} \\
V_{\text{V}}\left( z\right) &=&\frac{3}{2}\left[ -A^{\prime \prime }+\frac{3%
}{2}A^{\prime 2}-3\frac{A^{\prime }}{z}+\frac{1}{2}\frac{1}{z^{2}}\right]
+m_{5,V}^{2}R^{2}\frac{e^{2A}}{z^{2}}  \label{vva}
\end{eqnarray}%
as well as 
\begin{equation}
V_{\text{B},\pm }\left( z\right) =m_{5,\text{B}}R\frac{e^{A}}{z}\left[ \pm
\left( A^{\prime }-\frac{1}{z}\right) +m_{5,\text{B}}R\frac{e^{A}}{z}\right]
.  \label{vba}
\end{equation}%
The AdS/CFT boundary condition for the bulk modes at small $z$, which
relates them to the twist dimensions $\bar{\tau}_{\text{M}}=L+2$, $\bar{\tau}%
_{B}=L+3 $ of the dual hadron interpolating operators, is then imposed by
adjusting the bulk mode masses as \cite{for07} 
\begin{eqnarray}
m_{5,\text{S}}^{2}R^{2} &=&\bar{\tau}_{\text{M}}(\bar{\tau}_{\text{M}%
}-4)=L^{2}-4,  \label{m2m} \\
m_{5,\text{V}}^{2}R^{2} &=&\bar{\tau}_{\text{M}}(\bar{\tau}_{\text{M}%
}-4)+3=L^{2}-1,  \label{m2v} \\
m_{5,\text{B}}R &=&\bar{\tau}_{\text{B}}-2=L+1.  \label{mb}
\end{eqnarray}%
The lightest string modes are associated with the leading twist operators,
and therefore with the valence quark content of the low-spin (i.e. spin 0,
1/2, 1, and 3/2) hadron states \cite{hwph,bro04}. The duals of their orbital
excitations (which have no counterparts in the supergravity spectra) are
identified with fluctuations about the AdS background \cite{hwph,bro04}.
(This identification is incomplete, however, as long as quark flavor is not
explicitly accounted for.)

In order to search for IR deformations $A\left( z\right) $ which generate
the linear trajectorial (LT) structure (\ref{nltraj}), we first determine
the required potentials $V_{\text{M},\text{B}}^{\left( \text{LT}\right) }$.
They should be rising quadratically with $z$ for $%
z\rightarrow \infty $ to yield an equidistant spectrum for the higher-lying
excitations. The more challenging question is how to obtain a universal
slope $W$ in both meson and baryon channels. It turns out that this can be
achieved by replacing $\bar{\tau}_{i}\rightarrow \bar{\tau}_{i}+\lambda
^{2}z^{2}$ in the pure AdS$_{5}$ potentials (i.e. Eqs. (\ref{vma}) - (\ref%
{vba}) with $A\equiv 0$) \cite{for07}, leading to%
\begin{equation}
V_{\text{M}}^{\left( \text{LT}\right) }\left( z\right) =\left[ \left(
\lambda ^{2}z^{2}+L\right) ^{2}-\frac{1}{4}\right] \frac{1}{z^{2}}
\label{vMconf}
\end{equation}%
(which holds for both spin 0 and 1) and 
\begin{equation}
V_{\text{B},\pm }^{\left( \text{LT}\right) }\left( z\right) =\left\{ \left(
L+1\right) \left( L+1\mp 1\right) +\left[ 2\left( L+1\right) \pm 1\right]
\lambda ^{2}z^{2}+\lambda ^{4}z^{4}\right\} \frac{1}{z^{2}}.  \label{vBconf}
\end{equation}%
The normalizable solutions of the corresponding eigenvalue problems (\ref%
{eveqm}) and (\ref{eveqb}) can be found analytically \cite{for07}. The
eigenvalues 
\begin{equation}
M_{\text{M}}^{2}=4\lambda ^{2}\left( N+L+\frac{1}{2}\right) ,\text{ \ \ \ \ }%
M_{\text{B}}^{2}=4\lambda ^{2}\left( N+L+\frac{3}{2}\right)  \label{bspec}
\end{equation}
indeed generate the observed trajectories (\ref{nltraj}) with universal
slope $W=4\lambda ^{2}$ and a mass gap of order $\sqrt{W}$. They further
imply the new relations $M_{\text{M},0}^{2}=W/2$, $M_{\text{B},0}^{2}=3W/2$
between the ground state masses and the trajectory slope.

One has now to check whether the potentials (\ref{vMconf}), (\ref{vBconf})
can emerge from stringy fluctuations in a bulk gravity background 
(\ref{metric}). We do
this by construction, i.e. by equating the general-$A$ potentials (\ref{vma}%
) - (\ref{vba}) to their heuristic counterparts (\ref{vMconf}), (\ref{vBconf}%
), and by then searching for solutions of the resulting differential
equations for $A\left( z\right) $ subject to the physical boundary
conditions. \emph{A priori }the existence of such a bulk geometry is far
from guaranteed since the potentials (\ref{vMconf}), (\ref{vBconf}) may not
result from a boundary gauge theory. This is reflected in the fact that the
nonlinear, inhomogeneous differential equations for $A$ may not have
physically acceptable solutions.

\begin{figure}[tbp]
\includegraphics[width=.5\textwidth]{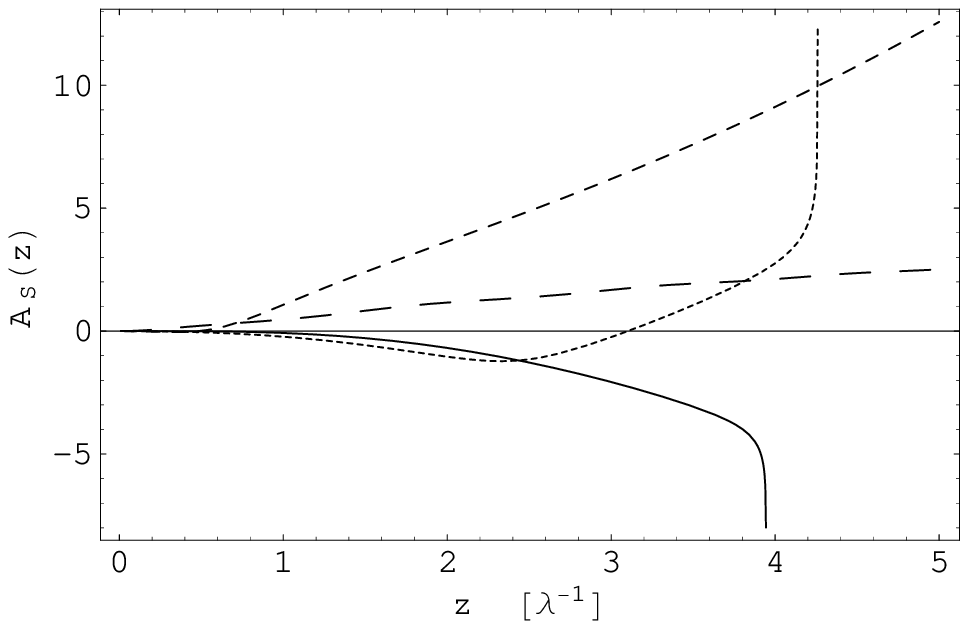} \includegraphics[width=.5%
\textwidth]{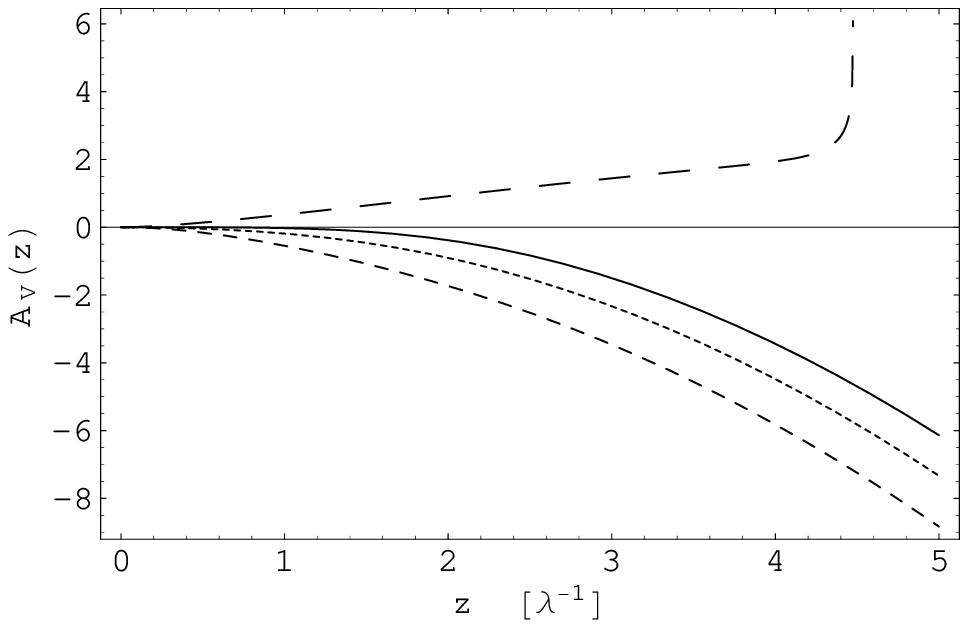}
\caption{ Typical solutions $A_{S}\left( z\right) $ (left panel) and $%
A_{V}\left( z\right) $ (right panel) for $L=0$ (full line), $L=1$ (dotted
line), $L=2$ (short-dashed) and $L=3$ (long-dashed). The dual eigenmodes
have significant support only for $z<\protect\sqrt{2}\protect\lambda ^{-1}$.}
\label{am}
\end{figure}
As we have shown in Ref. \cite{for07}, however, physically sensible IR
deformations $A\left( z\right) $ indeed emerge as unique solutions. In the
baryon sector, the solution $A_{\text{B}}\left( z\right) $ subject to the
conformal boundary condition $A_{\text{B}}\left( 0\right) =0$ can be found
analytically (for both chiralities), 
\begin{equation}
A_{\text{B}}\left( z\right) =\ln \left( 1+\frac{\lambda ^{2}z^{2}}{m_{5,%
\text{B}}R}\right) =\ln \left( 1+\frac{\lambda ^{2}z^{2}}{L+1}\right) .
\label{bsoln}
\end{equation}%
The analogous solutions for $A_{\text{S}}$ in the spin-0 meson and $A_{\text{%
V}}$ in the vector meson channel, which were numerically obtained and
discussed in Ref. \cite{for07}, are plotted for $L=0,...,3$ in Fig. \ref{am}%
. The small-$z$ behavior of these solutions hints at the formation of a
two-dimensional condensate and indicates its relevance for linear confinement. 
The $L$ dependence of the resulting $A\left( z\right) $ may be interpreted as
describing $L$ dependent stringy quantum fluctuations about the AdS
background which deform its metric in an $L$ dependent fashion \cite{for07}.
The nature of the singularities in several mesonic IR deformations, their
relation to the RG flow of the associated QCD interpolators, and possible
relations to the color-dielectric QCD vacuum structure are also discussed in
Ref. \cite{for07}.

\begin{figure}[tbp]
\includegraphics[width=.5\textwidth]{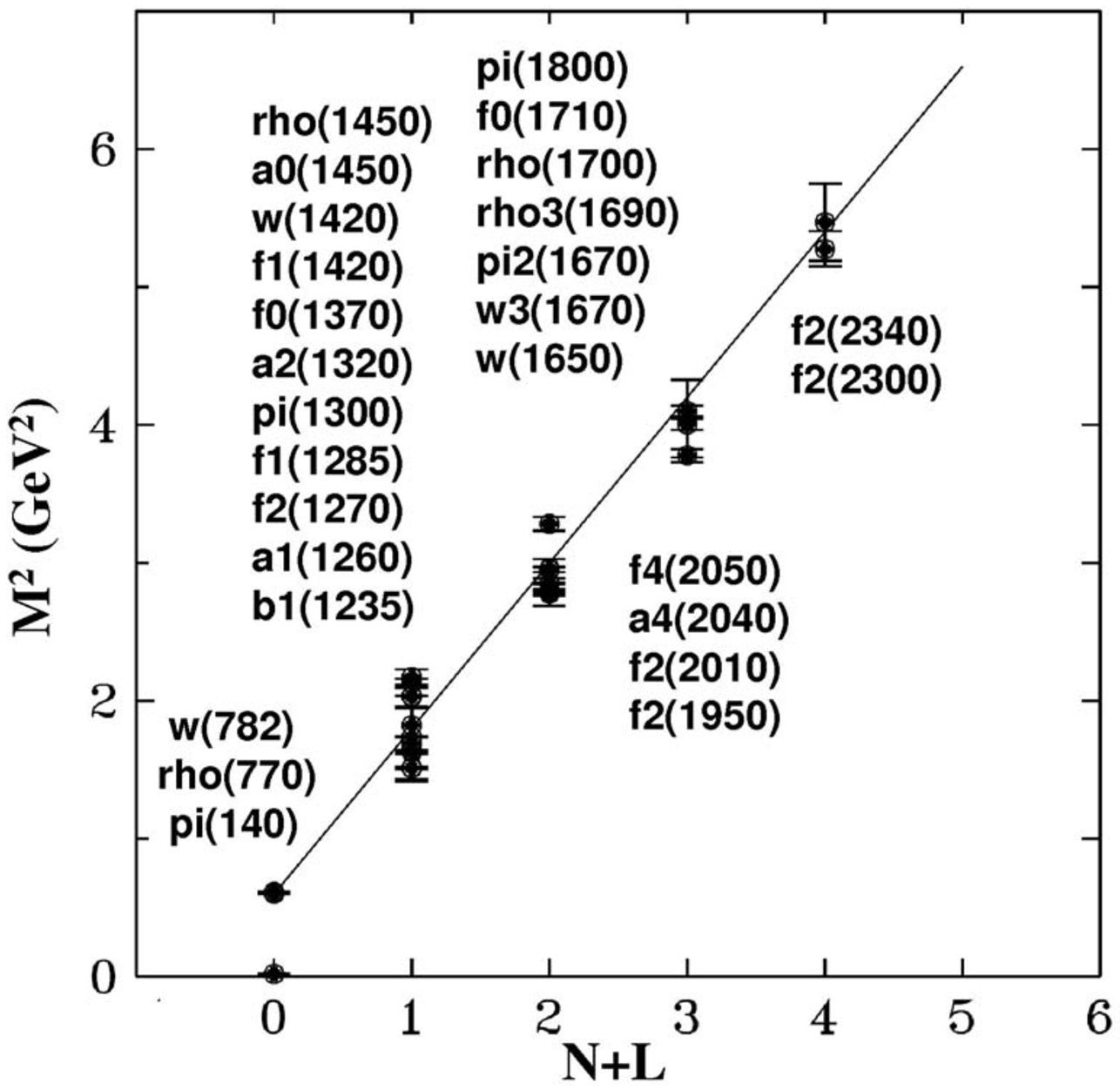} \includegraphics[width=.5%
\textwidth]{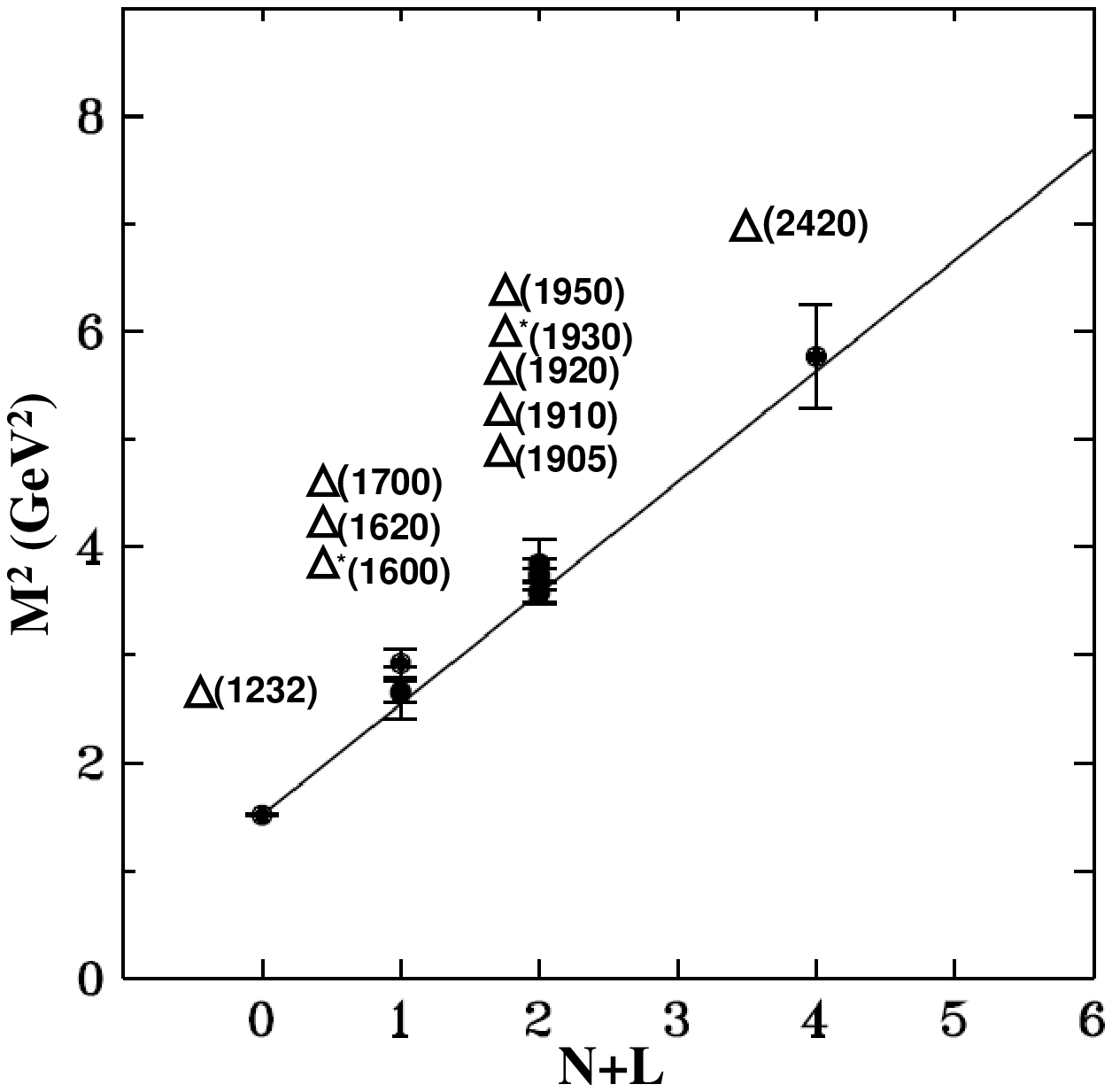}
\caption{Left panel: Experimental meson mass spectrum from Ref. \protect\cite%
{pdg} and the predicted trajectory for $W=2M_{{\protect\rho}}^{2}\simeq
1.21 $ GeV$^{2}$; Right panel: same for the Delta isobar mass spectrum with $%
S=3/2 $ (in the $^{4}$8 representation of SU(4)) and with $W=2M_{\Delta
}^{2}/3\simeq 1.01$ GeV$^{2}$. }
\label{fig1}
\end{figure}

The resulting, overall description of the excited hadron spectra \cite{jaf06}
is surprisingly accurate \cite{for07}. Using the experimental rho meson mass $%
M_{\rho }=0.76$ GeV~\cite{pdg} to set the deformation scale $\lambda $, the
resulting slope $W=1.21$ $\mathrm{GeV}^{2}$ and spectrum reproduce the
experimental meson masses (for quark-antiquark states), as shown in the left
panel of Fig.~\ref{fig1}. 
(The pion ground state does not fit into the overall pattern, due to the
lack of explicit chiral symmetry.) Alternatively, we can use the
experimental $M_{\Delta }$ to determine $\lambda =0.50\ \mathrm{GeV}$ which
differs by less than 10\% from the value in the meson sector and reflects
the approximate slope universality (cf. right panel of Fig.~\ref{fig1}). The
nucleon excitations require a somewhat smaller value $\lambda =0.47\ \mathrm{%
GeV}$ and are less well described by the trajectory (\ref{bspec}) which
overestimates, in particular, the ground-state mass (yielding $M_{\text{N}}=1.16\ 
\mathrm{GeV}$). Finally, the estimates $\Lambda _{\mathrm{QCD}}\simeq \sqrt{%
W/8}\simeq 0.35~$GeV of the QCD scale and $\sigma =W/\left( 2\pi \right)
\simeq 0.88$ GeV/fm for the string tension turn out to be close to the
empirical values.

\section{Holographic diquark correlations and the nucleon excitation spectrum%
}

\label{hdc}

While the metric-soft-wall prediction (\ref{bspec}) works remarkably well in
the $\Delta $ sector \cite{kle08} (all observed $\Delta ^{\ast }$ resonance
states lie within errors on the predicted trajectory with empirical slope
corresponding to $\lambda =0.52$ GeV), the description of the nucleon data
is poorer \cite{for07}. In the following section we review our recent
extension \cite{for09} of the metric-soft-wall dual which generates a
universal additive correction%
\begin{equation}
\Delta M_{\text{B},\kappa _{\text{gd}}}^{2}=-2\left( M_{\Delta
}^{2}-M_{N}^{2}\right) \kappa _{\text{gd}}  \label{dcmf}
\end{equation}%
to Eq. (\ref{bspec}) which solely depends on the resonances' diquark
content. The latter enters through the good (i.e. most attractive)\ diquark
fraction $\kappa _{\text{gd}}$ in the space-spin-flavor baryon wavefunction
(i.e. $\kappa _{\text{gd}}$\thinspace =\thinspace 0 for all $\Delta $ and
spin-3/2 $N$ resonances, $\kappa _{\text{gd}}$\thinspace =\thinspace 1/4 for
the spin-1/2 negative-parity $N$ resonances, and $\kappa _{\text{gd}}$%
\thinspace =\thinspace 1/2 for nucleons in the ground state). In order to
compare Eqs. (\ref{bspec}), (\ref{dcmf}) to experimental data, one needs to
assign intrinsic orbital and spin angular momenta $L$ and $S$ to the
observed states. This has been done on the basis of quark model arguments
and extensively discussed in Ref. \cite{for09}. The correction (\ref{dcmf})
decidedly improves the agreement with all 48 measured nucleon and $\Delta$
masses, beyond any dynamical quark model prediction of the full mass
spectrum.

In searching for a\ transparent origin of the universal mass correction (\ref%
{dcmf}) in AdS/QCD, one is led to ask how the diquark content of the baryon
resonances can enter a holographic description although diquarks and their
operators are gauge dependent while only gauge-invariant operators have
well-defined dual modes. The answer to this pivotal question lies in the
(leading-twist) baryon interpolating fields \cite{esp83} 
\begin{equation}
\eta _{t}\left( x\right) =2\left[ \eta _{\text{pd}}\left( x\right) +t\eta _{%
\text{sd}}\left( x\right) \right]  \label{ni}
\end{equation}%
of QCD which contain gauge-invariant diquark information through the
pseudoscalar diquark operator in $\eta _{\text{pd}}=\varepsilon _{abc}\left(
u_{a}^{T}Cd_{b}\right) \gamma ^{5}u_{c}$ and the \textquotedblleft
good\textquotedblright\ scalar diquark operator in $\eta _{\text{sd}%
}=\varepsilon _{abc}\left( u_{a}^{T}C\gamma ^{5}d_{b}\right) u_{c}$. (Here
we specialize to $N_{c}=3$, as elsewhere in AdS/QCD.) The interpolators (\ref%
{ni}) are expected to have enhanced overlap with nucleon states of
equivalent diquark content and are thereby associated with their
good-diquark fraction. This manifests itself in $\kappa _{\text{gd}}$
dependent anomalous dimensions $\gamma _{t^{\left( \kappa _{\text{gd}%
}\right) }}$ of the corresponding interpolators $\eta _{t^{\left( \kappa _{%
\text{gd}}\right) }}$ which holographically induce mass corrections $\Delta
m_{5}^{\left( \kappa _{\text{gd}}\right) }=\gamma _{t^{\left( \kappa _{\text{%
gd}}\right) }}$ for the dual modes \cite{revs} (as they could also arise
e.g. from $\kappa _{\text{gd}}$ dependent couplings of the dual modes to
other bulk fields).

In order to include the contributions from the so far neglected anomalous
dimensions, we extend the metric soft wall by implementing
three bulk spinor fields $\Psi ^{\left( \kappa _{\text{gd}}\right) }$ (cf.
Eq. (\ref{5ddir})) dual to the interpolators $\eta _{t^{\left( \kappa _{%
\text{gd}}\right) }}$ with $\kappa _{\text{gd}}=0,$ $1/4$ and $1/2$,
respectively. These fields are defined as the solutions of the 5d Dirac
equation with bulk masses%
\begin{equation}
m_{5}^{\left( \kappa _{\text{gd}}\right) }=m_{5}^{\left( \text{ms}\right)
}+\Delta m_{5}^{\left( \kappa _{\text{gd}}\right) }=\frac{L+\Delta
m_{5}^{\left( \kappa _{\text{gd}}\right) }R+1}{R},\text{ \ \ \ \ }\kappa _{%
\text{gd}}\in \left\{ 0,1/4,1/2\right\}  \label{m5}
\end{equation}%
which ensure that the chirally-odd components $\psi _{-}$ satisfy the
AdS/CFT boundary conditions. Together with the corresponding IR adjustment%
\begin{equation}
A_{B}\left( z\right) =\ln \left( 1+\frac{\lambda ^{2}z^{2}}{L+\Delta m_{5}R+1%
}\right)  \label{a}
\end{equation}%
of the warp factor (\ref{bsoln}), the corrected bulk masses (\ref{m5}) were
shown in Ref. \cite{for09} to generate a universal spectral correction of
the form (\ref{dcmf}). In the absence of reliable information on the
nonperturbative $\gamma _{t}$ we adjust%
\begin{equation}
\Delta m_{5}^{\left( \kappa _{\text{gd}}\right) }=\frac{\Delta M_{\kappa _{%
\text{gd}}}^{2}}{4\lambda ^{2}R}  \label{dm5}
\end{equation}%
in bottom-up fashion to reproduce the values of (\ref{dcmf}). The eigenvalue
spectrum (\ref{bspec}) then turns into the desired%
\begin{equation}
M_{N,L}^{2}=4\lambda ^{2}\left( N+L+\frac{3}{2}\right) -
2\left( M_{\Delta}^{2}-M_{N}^{2}\right) \kappa _{\text{gd}}.
\label{dcbms}
\end{equation}%
(The spectrum (\ref{dcbms}) can also be obtained when the RG flow of the
anomalous dimensions, which translates into a $z$ dependent $\Delta
m_{5}\left( z\right) $, is taken into account \cite{for09}.) Moreover, the
dual modes corresponding to larger $\kappa _{\text{gd}}$ feel the soft wall
at smaller $z$ and therefore extend less into the fifth dimension \cite%
{for09}. This reflects the additional attraction in the good-diquark channel
and translates into a smaller size of baryons with larger $\kappa _{\text{gd}%
}$.

\section{Dynamical AdS/QCD}

\label{dads}

The AdS/QCD duals discussed above share with most of those so far considered
the shortcoming that they are not solutions of a dual gravity. Hence the
dynamics which shapes the sought-after QCD dual remains obscure. Some of the
present dual candidates, including the dilaton soft-wall of Ref. \cite{kar06}%
, furthermore fail to exhibit the area-law behavior of the Wilson loop which
implies a linearly confining quark-antiquark potential \cite%
{dep09}. Others (including the hard wall (\ref{hw})) also confine magnetic
charges instead of screening them \cite{gue08}.

In Ref. \cite{dep09} we have shown how the above limitations can be
overcome, by deriving a confining AdS/QCD background from five-dimensional
Einstein-dilaton gravity 
\begin{equation}
S=\frac{1}{2\kappa ^{2}}\int d^{5}x\sqrt{\left\vert g\right\vert }\left( -%
\emph{R}+\frac{1}{2}g^{MN}\partial _{M}\Phi \partial _{N}\Phi -V(\Phi
)\right)  \label{actiongd}
\end{equation}%
with a metric restricted to the form (\ref{metric}) and a still general
potential $V$ for the dilaton $\Phi \left( z\right) $. More specifically, 
we search for\ static
solutions of the corresponding field equations for the background fields $A$
and $\Phi $, which we cast into the form%
\begin{equation}
\Phi ^{\prime }=\sqrt{3}\sqrt{-A^{\prime \prime }\left( z\right) +A^{\prime
2}\left( z\right) +\frac{2}{z}A^{\prime }\left( z\right) +\frac{2}{z^{2}}}
\end{equation}%
and%
\begin{equation}
V(\Phi \left( z\right) )=-\frac{3e^{-2A}}{2z^{2}}\left[ A^{\prime \prime
}\left( z\right) +3A^{\prime 2}\left( z\right) +\frac{6}{z}A^{\prime }\left(
z\right) +\frac{2}{z^{2}}\right] .
\end{equation}%
Our strategy is to construct solutions for the dilaton field $\Phi$ and 
potential $V(\Phi )$ after prescribing an area-law generating IR deformation 
$A$. More specifically, we adopt 
\begin{equation}
A(z)=-\frac{1+\sqrt{3}}{2S+\sqrt{3}-1}\frac{(z\Lambda _{\text{QCD}})^{2}}{%
1+e^{(1-z\Lambda _{\text{QCD}})}}  \label{cnew}
\end{equation}%
which generates a discrete spectrum with a mass gap and the area law while
keeping the fifth dimension non-compact to allow for linear Regge
trajectories. Eq. (\ref{cnew}) remains close to AdS$_{5}$ in the UV but
deforms rather rapidly for $z\gtrsim \Lambda _{\text{QCD}}^{-1}$ to approach
the confining large-$z$ asymptotics $A(z)\rightarrow z^{2}$. (The spin
dependent factor is required by universality. For a physical interpretation
see Ref. \cite{for07}.) We then find the corresponding dilaton field and 
potential numerically such that their combination solves the above
Einstein-dilaton equations.

\begin{figure}[tbh]
\includegraphics[width=.49\textwidth]{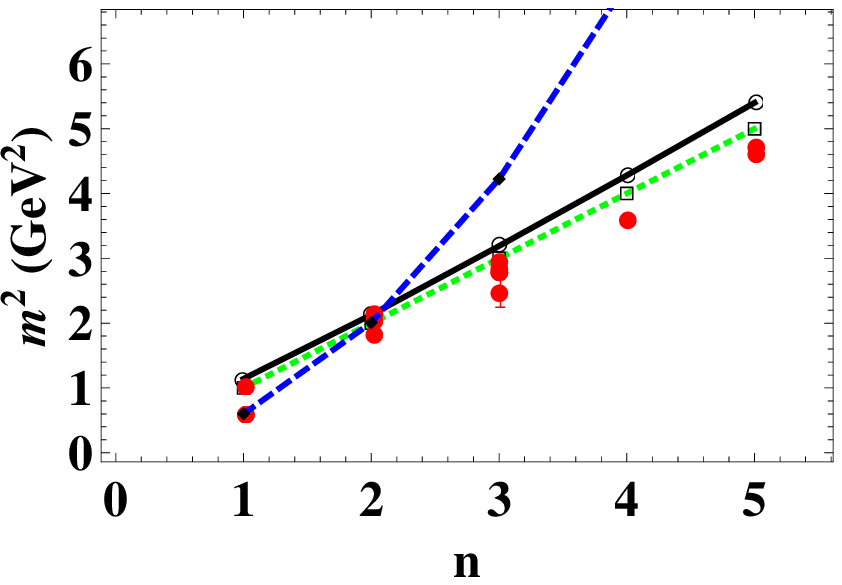} \includegraphics[width=.49%
\textwidth]{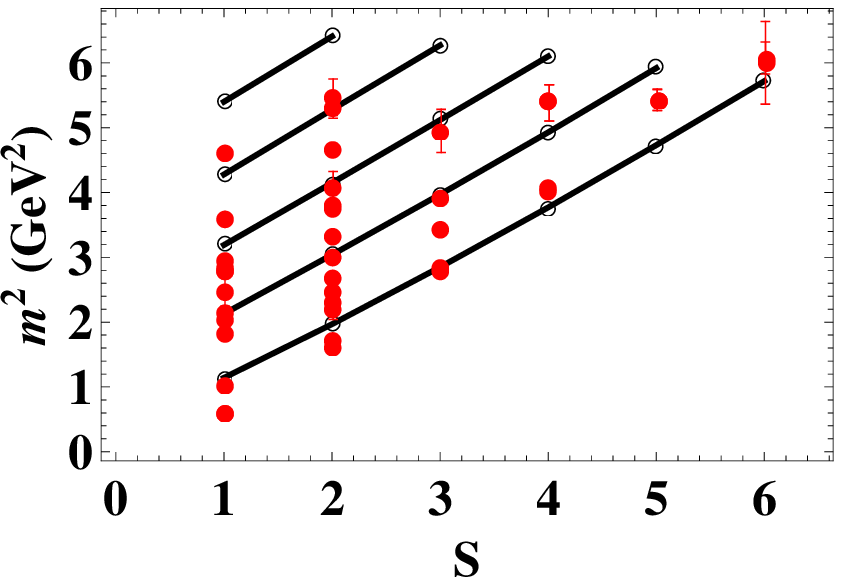}
\caption{(a) Radial excitations of the rho meson in the hard-wall (dashed
line), dilaton-soft-wall \protect\cite{kar06} (dotted line) and our dynamical
soft-wall (solid line, for $\Lambda _{\text{QCD}}=0.3$ GeV) backgrounds.
(Note that $n=1$ refers to the nodeless radial ground state.) (b) Square
mass predictions of spin excitations vs. the PDG values \protect\cite{pdg}.}
\label{fig4}
\end{figure}

The ansatz (\ref{cnew}) is furthermore designed to generate (approximately)
linear Regge trajectories in the highly excited meson spectrum. This
spectrum is derived in the tensor gauge-field framework of Ref. \cite{kar06}
which leads to a spin-dependent string-mode potential 
\begin{equation}
\mathcal{V}_{S}(z)=\frac{B^{\prime }{}^{2}(z)}{4}-\frac{B^{\prime \prime }(z)%
}{2}  \label{vs}
\end{equation}%
with $B=-\left( 2S-1\right) \left( \ln z+A\right) +\Phi $. Important
qualitative aspects of the meson spectrum can be understood by studying the
UV (i.e. $z\rightarrow 0$) and IR ($z\rightarrow \infty $) limits of the
mode potential \cite{dep09}. In Fig. (\ref{fig4}) the resulting spectrum is
compared to experimental data and hard- and dilaton-soft-wall predictions. A
satisfactory description of the meson spectrum with nearly linear
trajectories of universal slope is indeed achieved without tuning adjustable
parameters, as testified by the rather accurate parametrization 
\begin{equation}
m_{n,S}^{2}\simeq \frac{1}{10}\left( 11n+9S+2\right) ,\text{ \ \ \ \ \ }%
\left( n\geq 1\right)
\end{equation}
(in units of GeV, for $\Lambda _{\text{QCD}}=0.3$ GeV) which makes the
approximate slope universality explicit.

Asymptotic freedom and the perturbative corrections to it could
additionally be implemented into Eq. (\ref{cnew}) for small $z\ll \Lambda _{%
\text{QCD}}^{-1}$, according to the perturbative QCD $\beta $ function. This
generates a leading correction $A_{\text{pert}}(z)=(2\ln z)^{-1}${\ which
naturally coexists with confinement at large $z$.}

\section{Holographic glueball correlators}

In order to make progress with the construction of improved AdS/QCD duals, 
one eventually has to analyze more complex and detailed amplitudes. A natural 
choice are
correlation functions of hadronic interpolators which are both
directly accessible from the AdS/CFT dictionary and in several cases
well-studied in QCD. With this motivation in mind, we have recently 
derived and analyzed the
predicitions of two popular AdS/QCD duals, the hard-wall (\ref{hw}) and
dilaton soft-wall \cite{kar06} backgrounds, for the $0^{++}$ glueball
correlation function and decay constants \cite{for08}. Since this work was
reviewed in Ref. \cite{for208}, we will restrict ourselves here to a brief
summary. (For related work in the dilaton soft-wall at finite temperature
see Ref. \cite{mir09}.)

Both holographic duals turn out to complement each other in their
representation of specific nonperturbative glueball physics (at momenta
larger than the QCD scale): the soft-wall correlator 
\begin{eqnarray}
\hat{\Pi}^{\mathrm{(sw)}}\left( Q^{2}\right) &=&-\frac{2R^{3}}{\kappa ^{2}}
\lambda^{4}\left[ 1+\frac{Q^{2}}{4\lambda ^{2}}\left( 1+\frac{Q^{2}}{%
4\lambda ^{2}}\right) \psi \left( \frac{Q^{2}}{4\lambda ^{2}}\right) \right]
\notag \\
&&\overset{Q^{2}\gg \lambda ^{2}}{\longrightarrow }-\frac{2}{\pi ^{2}}Q^{4}%
\left[ \ln \frac{Q^{2}}{\mu ^{2}}+\frac{4\lambda ^{2}}{Q^{2}}\ln \frac{Q^{2}%
}{\mu ^{2}}+\frac{2^{2}5}{3}\frac{\lambda ^{4}}{Q^{4}}-\frac{2^{4}}{3}\frac{%
\lambda ^{6}}{Q^{6}}+\frac{2^{5}}{15}\frac{\lambda ^{8}}{Q^{8}}+...\right]
\label{psw}
\end{eqnarray}%
(where $\psi \left( z\right) =\Gamma ^{\prime }\left( z\right) /\Gamma
\left( z\right) $, $\lambda$ is the dilaton mass scale and $R^{3}/\kappa
^{2}=2(N_c^2-1)/\pi^2$) contains all known types of QCD power corrections,
generated both by vacuum condensates and by a hypothetical UV gluon mass
suggested to encode the short-distance behavior of the static
quark-antiquark potential \cite{che99}, while sizeable exponential
corrections as induced by small-scale QCD instantons \cite{for01} are
reproduced in the hard-wall correlator%
\begin{eqnarray}
\hat{\Pi}^{\mathrm{(hw)}}\left( Q^{2}\right) &=&\frac{R^{3}}{8\kappa ^{2}}%
Q^{4} \left[ 2\frac{K_{1}\left( Qz_{m}\right) }{I_{1}\left( Qz_{m}\right) }%
-\ln \frac{Q^{2}}{\mu ^{2}} \right]  \notag \\
&&\overset{Q^{2}\gg z_{m}^{-2}}{\longrightarrow }-\frac{2}{\pi ^{2}}Q^{4}\ln 
\frac{Q^{2}}{\mu ^{2}} +\frac{4}{\pi }\left[ 1+\frac{3}{4}\frac{1}{Qz_{m}}%
+O\left( \frac{1}{\left( Qz_{m}\right) ^{2}}\right) \right] Q^{4}e^{-2Qz_{m}}
\label{phw}
\end{eqnarray}%
(where the IR brane is located at $z_m$). This complementarity generalizes
to other hadron channels, allows to relate holographic predictions to
specific aspects of the gauge dynamics and suggests to combine the
underlying brane- and dilaton-induced IR physics into improved QCD duals.

While the various contributions to the holographic estimates (\ref{psw}) and
(\ref{phw}) have the expected order of magnitude, the signs of the two
leading power corrections in Eq. (\ref{psw}) are opposite to QCD predictions
and violate the factorization approximation to the four-gluon condensate. We
have argued that this provides specific evidence for the short-distance
physics in the Wilson coefficients to be inadequately reproduced by the
strongly-coupled UV dynamics of the gravity duals (beyond the leading
conformal logarithm) \cite{for08}. (This problem cannot be mended by
admixing the UV-subleading solution to the bulk-to-boundary propagator \cite%
{col07} without loosing consistency and predictive power \cite{for208}.) It
remains to be seen whether $\alpha ^{\prime }$ corrections, in particular
the resummed, local ones which are suggested to reproduce the RG flow of the
gauge coupling \cite{gue08}, can generate improved holographic predictions
for the power corrections.

Since the QCD Wilson coefficients of the $0^{++}$ glueball correlator receive
unusually small perturbative and enhanced instanton contributions, the
hard-wall correlator may yield the better overall AdS/QCD description. Our
holographic estimates of the glueball decay constants, which are important
for experimental glueball searches, provide further evidence for this
expectation. The large hard-wall prediction $f_{S}^{\left( \text{hw}\right)
}\simeq 0.8-0.9$ GeV for the ground-state decay constant, in particular, 
reflects the strong
instanton-induced short-distance attraction in the scalar QCD glueball
correlator, implies an exceptionally small $0^{++}$ glueball size and is
indeed close to IOPE sum-rule \cite{for01} and lattice \cite{che06} results.
The absence of instanton contributions in the soft wall with its
confinement-induced linear meson trajectories, on the other hand, may
suggest that instantons are not directly involved in flux-tube formation.

\section{Summary and conclusions}

\label{sum}

We have reviewed our recent work on four topics in holographic QCD. To begin
with, we have sketched the construction of the \textquotedblleft metric soft
wall\textquotedblright\ dual and discussed its capacity to reproduce the
empirical combination of radially and orbitally excited hadron mass
spectra into linear trajectories of approximately universal slope. The
resulting bulk background is solely based on IR\ deformations of the AdS
metric, encodes dual signatures of linear quark confinement and contains
only one adjustable parameter related to the string tension. It so far 
remains the only AdS/QCD dual which is able to reproduce linear trajectories 
also in the baryon sector. The predicted spectra, as well as new relations
between the $\rho $ and $\Delta $ ground state masses and the slopes of
their respective trajectories, are in good overall agreement with
experimental data.

The metric-soft-wall predictions for the nucleon and its excitations turn
out to be significantly less accurate than those in the meson and $\Delta $
sectors, however. This led us to extend this dual by holographically
encoding the diquark content of the light baryon states. The latter is
specified by the good-diquark fraction of the corresponding baryon
interpolators whose anomalous dimensions are then translated by the AdS/CFT
dictionary into dual string-mode mass corrections. After implementing the
diquark correlations, the improved metric soft wall reproduces the masses of
all 48 observed nucleon and $\Delta $ resonances with far better accuracy
than e.g. quark models based on substantially larger parameter sets. The
behavior of the corresponding bulk modes further reveals that the sizes of
the light-quark baryons decrease when their good-diquark content increases.

Another focus of our work was the search for higher-dimensional gravitational
dynamics which are capable of generating approximate holographic QCD backgrounds. 
In particular, we have derived a new solution of the
five-dimensional Einstein-dilaton equations with a specific dilaton potential
which generates a confining area
law for the Wilson loop and can implement the perturbative running of the
gauge coupling. It further encodes linear square-mass trajectories for 
both radial and spin excitations in the meson sector and reproduces the
approximately universal slope of the observed trajectories. The result is 
a satisfactory, fully dynamical description of the light-flavored
natural-parity meson spectrum without adjustable parameters beyond the QCD
scale.

In order to study the holographic dynamics in more detail, we have further 
derived and
analyzed the $0^{++}$ glueball correlation function and its spectral density
in the hard-wall and dilaton soft-wall gravity duals. The resulting
expressions were confronted with QCD information from the lattice, the
instanton-improved operator product expansion (OPE), low-energy theorems
etc. This analysis revealed, in particular, that the soft-wall correlator
contains all known types of QCD power corrections (including those generated
by an effective UV gluon mass) while the hard-wall correlator exhibits in a
complementary fashion large exponential corrections as induced by
small-scale instantons. The results further show that the comparison of
holographic predictions with QCD results at the correlator level
provides valuable diagnostic insights into the limitations of the underlying
duals and leads to useful suggestions for their improvement. The holographic
estimates of OPE Wilson coefficients, in particular, were shown to yield
detailed and quantitative information on the extent to which the underlying
short-distance physics is contaminated by the strongly-coupled UV regime of
bottom-up duals. We have further derived predictions for the glueball decay
constants which contain crucial size information and are of direct
importance for experimental glueball searches. Remarkably, the strong
instanton-induced attraction in the $0^{++}$ glueball channel is captured by
the hard-wall dual, and its prediction $f_{S}^{\left( \text{hw}\right)
}\simeq 0.8-0.9$ GeV for the ground-state decay constant agrees inside
errors with instanton-improved sum rule and lattice results.

Several current limitations notwithstanding, we conclude that the 
amount of QCD dynamics encoded in even the simplest holographic duals is 
encouraging and indicates that the bottom-up approach may eventually
turn into a systematic approximation for QCD. 

It is a pleasure to thank 
Tobias Frederico, Eberhard Klempt, Wayne de Paula
and Michael Beyer for their collaboration on different parts of the work
reviewed above, and the organizers and participants of QCD-TNT 2009 for a 
very informative and enjoyable workshop. Financial support from the 
Funda\c{c}\~{a}o de
Amparo \`{a} Pesquisa do Estado de S\~{a}o Paulo (FAPESP) and the Deutsche
Forschungsgemeinschaft (DFG) is also acknowledged.


\end{document}